\documentclass{pasa}

\title[Measuring Noise Temperatures of Array Antennas]{Measuring Noise Temperatures of Phased-Array Antennas for Astronomy at {CSIRO}}
\author[A. P. Chippendale, D. B. Hayman and S. G. Hay]{A. P. Chippendale$^{1,3}$, D. B. Hayman$^2$ \and S. G. Hay$^{2}$\\
\affil{$^1$CSIRO Astronomy and Space Science, PO Box 76, Epping, NSW 1710, Australia}%
\affil{$^2$CSIRO Computational Informatics, PO Box 76, Epping, NSW 1710, Australia}
\affil{$^3$Email: Aaron.Chippendale@csiro.au}}%
\jid{PASA}
\doi{10.1017/pas.\the\year.xxx}
\jyear{\the\year}
\usepackage[authoryear]{natbib}
\bibpunct{(}{)}{;}{a}{}{,}
\usepackage{graphicx}
\usepackage{aastex_hack}
\usepackage{amsmath}
\usepackage{algorithm}
\usepackage{algpseudocode}
\usepackage{textcomp}
\usepackage{url}

\newcounter{MYtempeqncnt}
\begin{document}
%
%
%
\begin{abstract}
We describe the development of a noise-temperature testing capability for phased-array antennas operating in receive mode from 0.7~GHz to 1.8~GHz. Sampled voltages from each array port were recorded digitally as the zenith-pointing array under test was presented with three scenes: (1) a large microwave absorber at ambient temperature, (2) the unobstructed radio sky, and (3) broadband noise transmitted from a reference antenna centred over and pointed at the array under test.  The recorded voltages were processed in software to calculate the beam equivalent noise temperature for a maximum signal-to-noise ratio beam steered at the zenith.  We introduced the reference-antenna measurement to make noise measurements with reproducible, well-defined beams directed at the zenith and thereby at the centre of the absorber target.  We applied a detailed model of cosmic and atmospheric contributions to the radio sky emission that we used as a noise-temperature reference.  We also present a comprehensive analysis of measurement uncertainty including random and systematic effects.  The key systematic effect was due to uncertainty in the beamformed antenna pattern and how efficiently it illuminates the absorber load.  We achieved a combined uncertainty as low as 4~K for a 40~K measurement of beam equivalent noise temperature.  The measurement and analysis techniques described in this paper were pursued to support noise-performance verification of prototype phased-array feeds for the Australian Square Kilometre Array Pathfinder telescope.
\end{abstract}



\maketitle%

\section{INTRODUCTION}
Developing low-noise, wideband, receive-only array antennas is crucial to delivering the Square Kilometre Array (SKA) telescope \citep{Dewdney2009}.  Using aperture arrays and phased-array feeds (PAFs) allows more information to be collected from more of the sky in parallel.  This increases instantaneous field of view, increases survey speed, and allows more agile observing strategies as electronic beam steering can be immediate.  

Array antennas enable telescope designers to spend more money on digital signal processing and less on mechanical signal processing via telescope dishes for a fixed performance goal.  This trade-off becomes more effective with time because digital signal-processing is becoming exponentially cheaper while the cost of dishes is not.  The SKA project explored this trade-off \citep{Schilizzi2007,Chippendale2007,Alexander2007} and settled on significant deployments of both PAFs and aperture arrays in SKA phase 1 \citep{Dewdney2013}. 


\begin{figure}
\begin{center}
\includegraphics[width=\columnwidth, angle=0]{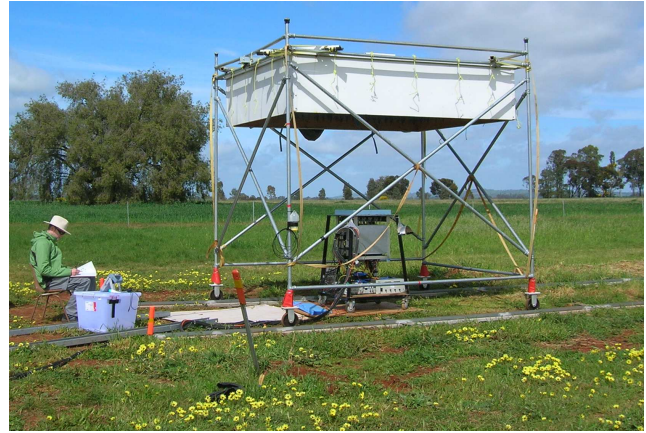}
\caption{Absorber rolled over array under test at Parkes.}\label{fig:hot-load}
\end{center}
\end{figure}

Important to the development of low-noise array antennas is the ability to make accurate and reproducible measurements of their noise performance after beamforming.  A common approach for measuring array noise temperature is to apply the same Y-factor method used for single-antenna astronomy receivers \citep{Sinclair1991} to the beamformed power from an array antenna \citep{Woestenburg2003}.  The Y-factor is the ratio of beamformed power between observations of ``hot'' and ``cold'' loads.  At decimetre wavelengths, the hot load is often provided by microwave absorber at ambient temperature (Figure \ref{fig:hot-load}) and the cold load by cosmic radio emission from the unobstructed sky.  

A number of groups have reported recent developments in test facilities and measurement techniques for low-noise arrays.  Some experiments have positioned the absorber at the end of a tapered metal funnel or ground shield \citep{Warnick2009,Woestenburg2011}. Others have used more open structures, like CSIRO's in Figure \ref{fig:hot-load}, to support the absorber over the array under test \citep{Woestenburg2012}. 

Previous work with a ground shield has indicated small differences between measured system noise temperatures with and without the shield \citep{Woestenburg2011}. The differences generally decrease with increase in the beamformed directivity of the array under test. At low directivity, where the shield is significant, the shield only partly decreased the effect of the terrestrial environment.  

\begin{figure}
\begin{center}
\includegraphics[width=\columnwidth, angle=0]{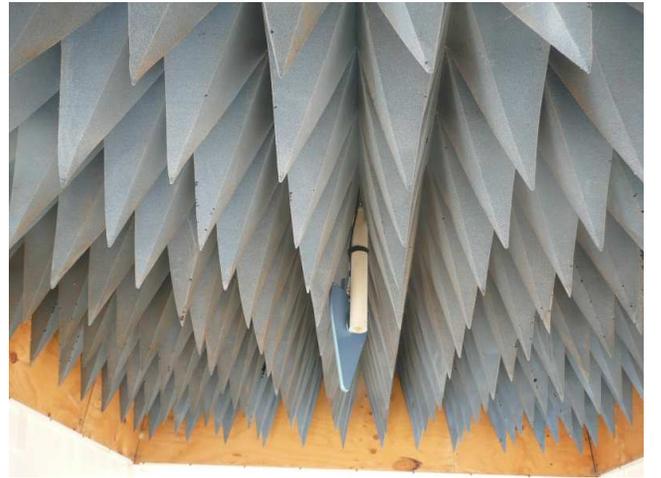}
\caption{Pyramidal foam absorber with log-periodic dipole array (LPDA) reference antenna located at centre.  The absorber is housed in a metal-backed wooden box.}\label{fig:absorber}
\end{center}
\end{figure}

In this paper we describe CSIRO's development of an aperture-array noise-temperature testing capability at Parkes Observatory.   We develop a Y-factor approach similar to \citep{Woestenburg2012} but introduce a reference-antenna (Figure \ref{fig:absorber}) measurement to constrain the pointing of the beam towards the zenith and therefore the centre of the absorber.

\section{PARKES TEST FACILITY}

The aperture-array test facility at CSIRO Parkes Observatory (-32$^\circ$59'56"S, 148$^\circ$16'3"E) uses a large rectangular microwave absorber supported by an open frame.  Figure \ref{fig:hot-load} shows that this absorber may be easily rolled over or away from the array under test via a wheel-on-track arrangement.  The aperture-array test pad is serviced by power, radio-frequency (RF) cabling, and a digital receiver and beamformer in a neighbouring hut.  


A nearby 12~m parabolic reflector has been used to test arrays at its focal plane using the same digital receiver as the aperture-array measurements.  Correlated measurements against signals from the 64~m Parkes radio telescope have also been used to boost testing capability in signal-to-noise ratio and the ability to measure phase \citep{Chippendale2010}.  The 64~m dish is located approximately 400~m west of the 12~m dish and aperture array test pad.

\section{ARRAY UNDER TEST}

We illustrate our aperture-array noise measurement procedure with real data from a prototype ${5 \times 4}$ element  ${\times \  2}$ polarisation (40-port) ``chequerboard'' connected element array \citep{Hay2008,Hay2011,Hay2007} hereafter called the ${5 \times 4}$ prototype.  This prototype was part of developing the larger 188-port phased-array-feed receivers \citep{Hay2010a, Hay2010, Schinckel2011, Hampson2012} for the Australian Square Kilometre Array Pathfinder (ASKAP) telescope \citep{DeBoer2009}.    

\begin{figure*}
\begin{center}
\includegraphics[scale=0.7, angle=0]{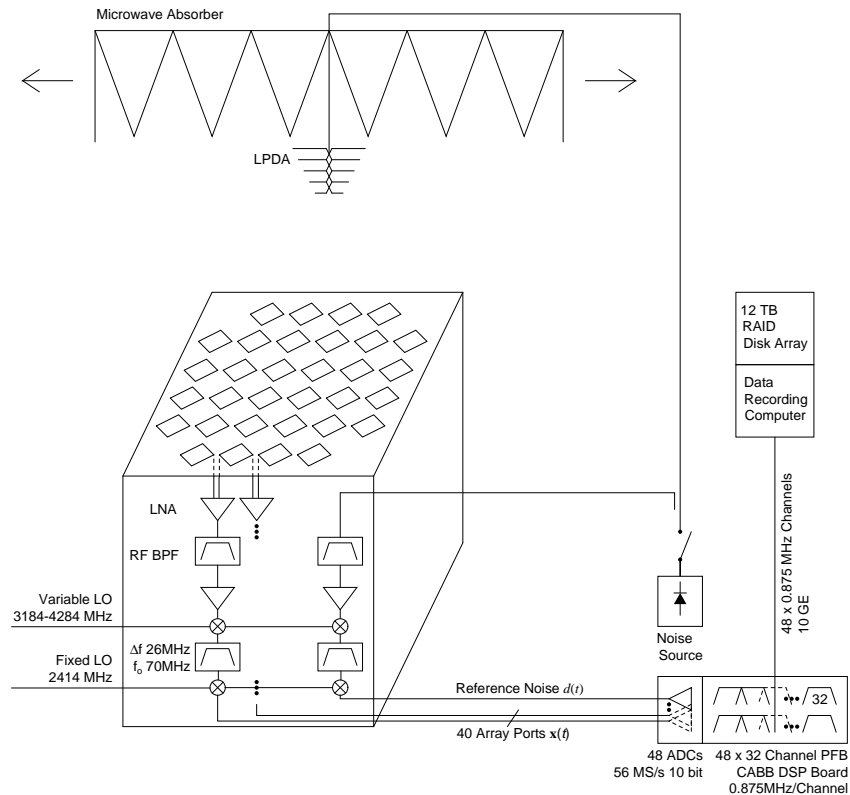}
\caption{Block diagram of beamformed noise performance measurement setup for a ${5\times4}$ prototype phased-array antenna.}\label{fig:schematic5x4}
\end{center}
\end{figure*}

\section{MEASUREMENT SYSTEM}

Noise measurements of the ${5 \times 4}$ prototype were made with a purpose built 48-port dual conversion superheterodyne receiver followed by a 48-port digitiser and field-programmable gate array (FPGA) based signal processor.  This initial measurement system was based on the same generation of technology as the New Technology Demonstrator \citep{Hayman2008,Hayman2010}.  The test facility has since been updated to use the same hardware that is deployed on the first six ASKAP antennas that form the Boolardy Engineering Test Array (BETA) \citep{Schinckel2011, Bunton2011}. 

Figure \ref{fig:schematic5x4} shows the measurement configuration for this paper.  Forty ports of the receiver were connected to the prototype array.  One of the spare receiver ports was connected to a directly coupled sample of the radiated noise source used to constrain beam direction.  The system was used to record baseband voltages with 0.875 MHz bandwidth to disk for each of the receiver's 48 ports.  Each 0.5~s packet of data was time-stamped with a precise measure of Universal Coordinated Time (UTC) from an atomic clock reference.  Although the system is capable of online beamforming, offline beamforming on recorded data allowed exploration of different beamforming and radio-frequency interference (RFI) removal strategies. 

Each LNA output was filtered, amplified, up-converted to an intermediate frequency (IF) of 2.484~GHz, and then down-converted to an IF of 70~MHz.  The 26~MHz bandwidth IF at 70~MHz was sampled at 56 MSPS then separated into $32 \times 0.875$~MHz channels by a digital polyphase filter bank (PFB) implemented in an FPGA based digital signal processing board\footnote{Compact Array Broadband (CABB) board} \citep{Wilson2011}.  The complex (I/Q) output of a single 0.875~MHz channel, fractionally oversampled by 8/7, was streamed via 10~Gbit Ethernet to a data recording computer attached to a RAID disk storage array.  The data recorder stored 0.5~s of contiguous I/Q data for each capture and was capable of approximately one capture every three seconds.  Oversampling by 8/7 meant that the sampling period was 1~{\textmu}s for the 0.875~MHz channel.

The absorber load, shown in Figure \ref{fig:hot-load}, is a 2,440~mm~$\times$~2,900~mm sheet of 610~mm (24~in) pyramidal foam absorber.  The manufacturer quotes a normal incidence reflectivity of $-40$~dB at 1~GHz. The foam is mounted tips-down in an upside-down sheet-metal box as shown in Figure \ref{fig:absorber}.  This mounting located the tips of the pyramids 2,590~mm above the ground and 1,270~mm above the surface of the array under test.  As the metal sides of the box come to just below the array tips, the effective height of the load above the array for calculating the region of sky blocked by the load is approximately 1,200~mm.

A log-periodic dipole array antenna (LPDA) is located at the centre of the absorber load as shown in Figure \ref{fig:absorber}. This is for radiating broadband noise into the array under test so that a beam may be steered towards the centre of the absorber load in a reproducible manner.  This antenna (Aaronia HyperLOG 7025) has a typical gain of 4~dBi from 0.7~GHz to 2.5~GHz.  The radiation patterns published by the manufacturer indicate that the illumination falls by approximately 0.3~dB from the centre to the edge of the array under test.

\section{MEASUREMENT OVERVIEW}

We adapted the Y-factor method to measure the equivalent noise temperature of a receive-only beamformed antenna array.  Over the 0.7~GHz to 1.8 GHz measurement band, the background radio sky has a ``cold'' brightness temperature of approximately 5~K away from the galactic plane, compared to a ``hot'' microwave absorber at ambient temperature near 300~K.  We deduce the noise contribution of the array from the Y-factor power ratio between beamformed measurements of the ``hot'' absorber and ``cold'' sky scenes. 

Voltages were recorded consecutively for six measurement states:
\begin{enumerate}
  \item sky;
  \item absorber;
  \item absorber and radiated, broadband noise;
  \item sky;
  \item absorber; and
  \item sky.
\end{enumerate}
For each state, the RF measurement frequency was swept from 0.6~GHz to 1.9~GHz in 100~MHz steps by tuning the variable local oscillator (LO).  Three 0.5~s recordings were made at each frequency for each measurement state.  Measurements at each state were separated by approximately seven minutes.  This consisted of three minutes to sweep the measurement frequency and record voltages for a given state, and four minutes to move the absorber and/or toggle the radiated noise source in preparation for recording the next state.

The measurements spanned the local Australian Eastern Standard Time (AEST) range at Parkes from 14:35~AEST to 15:11~AEST, which corresponded to the local sidereal time (LST) range from 17:21~LST to 17:57~LST.  The centre of this time range 17:39~LST (14:53 AEST) corresponded closely to the transit of the galactic centre which occurs at 17:45~LST.  In fact, 17:39~LST corresponds exactly to the epoch at which maximum antenna temperature is expected during a zenith drift-scan with a low-gain antenna from a latitude near 30$^\circ$S \citep{Chippendale2009}.  At the midpoint of observations the Sun was at azimuth 281.8$^\circ$ and elevation 44.9$^\circ$ and was therefore just blocked by the absorber when it was rolled over the array under test.  

The physical temperature of the absorber $T_\text{abs}$ was taken as the mean ambient temperature measured by the observatory's weather station over the measurement period.  This resulted in ${T_\text{abs} = 294.2 \pm 1 \mathrm{K}}$ where the uncertainty was estimated by the standard deviation of the temperature measurements. The air pressure used for atmospheric emissivity calculation was 973~hPa as measured by the same weather station.  The beamformed antenna temperature when observing the absorber was calculated by convolving the array factor pattern with the model sky brightness masked by an ideal model of the absorber with uniform brightness equal to its physical temperature.  Diffraction about the edges of the absorber and scattering from its supporting frame were not considered.


\section{BEAMFORMING METHOD}
We introduced a technique to ensure noise measurements were made with well defined and reproducible beams directed at the centre of the absorber.  Beam direction and polarisation were constrained by measurements of a radiated noise source located at the centre of the absorber as shown in Figure \ref{fig:absorber}.   

Beamforming was performed offline in software using maximum signal-to-noise ratio ($S/N$) weights \citep{lo1966}.  These were calculated by the method of direct matrix inversion developed by \cite{Reed1974} and summarised by \cite{Monzingo2011}.   

First, the receiver-output sample correlation matrix was calculated by 
\begin{equation}
  \mathbf{R}_{xx} = \frac{1}{L}\sum_{n=1}^{L}\mathbf{x}(n)\mathbf{x}^{H}(n)
  \label{eq:scm}
\end{equation}
where $\mathbf{x}(n)$ is the $n^{\text{th}}$ time sample of the column vector of 40 complex array-port voltages $\mathbf{x}(t)$.  Second, beamformed power $P$ for weight vector $\mathbf{w}$ was calculated by
\begin{equation}
  P = \mathbf{w}^H\mathbf{R}_{xx}\mathbf{w}.
\end{equation}

For this work we used maximum $S/N$ weights estimated via direct inversion of the sample noise correlation matrix $\mathbf{R}_{nn} $.  This noise correlation matrix is calculated according to \eqref{eq:scm} from data recorded when the array observed the unobstructed sky.  The maximum $S/N$ weights are given by \citep{lo1966, Widrow1967}
\begin{equation}
  \mathbf{w} =\mathbf{R}_{nn}^{-1}\mathbf{r}_{xd}
  \label{eq:maxsnrweights}
\end{equation}
where $\mathbf{r}_{xd} $ is the sample cross-correlation vector
\begin{equation}
  \mathbf{r}_{xd} = \frac{1}{L}\sum_{n=1}^{L}\mathbf{x}(n)d^{*}(n)
\end{equation}
and $d(t)$ is a reference signal provided as a template of the desired signal.

Figures \ref{fig:absorber} and \ref{fig:schematic5x4} show how we generated a reference signal by radiating broadband noise from an LPDA antenna located directly above the array.  The noise source was fed through a coupler so that a copy of the radiated noise could be recorded directly via a spare port of the receiver.  This allowed high $S/N$ measurement of $\mathbf{r}_{xd} $ while keeping the radiated noise source weak enough that it increased the noise power measured at individual array ports by just 3~dB.  The plane of polarisation of the LPDA was oriented at 45$^\circ$ to the plane of polarisation of the array elements.

The desired reference signal for well defined aperture-array noise measurements is a plane wave from boresight.  Although the radiator used as the source for beamforming is only 1.27~m from the array, the near-field effect is expected to be small.  Electromagnetic modelling of the experiment indicates less than 3~K variation in beamformed noise temperatures due to the near-field effect.


The maximum $S/N$ weights calculated from the sample cross-correlation $\mathbf{r}_{xd}$ with the reference antenna signal are in fact equivalent to least-mean-square (LMS) beamforming \citep{Widrow1967,Compton1988}.  The LMS algorithm minimises the square of the difference between the beamformed phased-array voltage and the directly-coupled copy of the broadband noise voltage transmitted from the reference radiator.

We used $L=500,000$ samples to calculate $\mathbf{R}_{nn}$ and $\mathbf{r}_{xd} $ for making weights via \eqref{eq:maxsnrweights} in each 0.875~MHz channel for beamformed noise measurements.  We also verified the convergence of these weights by inspecting plots of weight amplitude, phase, and beamformed noise temperature versus the number of samples $L$ used.  We believe this probes the convergence of $\mathbf{R}_{nn}$ as we measured $\mathbf{r}_{xd}$ with much higher $S/N$ due to correlation against the coupled copy of the reference noise.  Verifying convergence times against theory boosted our confidence that the measurement system operated as expected, and that the maximum $S/N$ weight solution was not being perturbed by gain fluctuations or non-stationary RFI. 

The measured noise temperature converged to within a factor of two of its minimum after 50 samples and to within 2\% of its minimum after 2,000 samples.  Both of these convergence checks agree well with the theoretical expectation for relative excess output residue power given by \citep{Monzingo2011,Reed1974} as 
\begin{equation}
  \left<r^2\right> = \frac{M}{L-M}.
\end{equation}
This predicts convergence to within a factor of two after $2M=80 $ samples and to within 2\% after $51M=2,040 $ samples where $M=40$ is the number of array ports.  

\section{DATA SELECTION}
Having observed that the weights converge sufficiently after 2,000 samples, we reduced all available data by calculating $\mathbf{R}_{xx} $ and $\mathbf{r}_{xd} $ with $L=2,000$ samples.  This generated ${250\times2}$~ms measurement points from each 0.5 s baseband data file.  Before further processing, each 2~ms measurement was analysed for positive outliers in total power that are expected due to transient radio-frequency interference (RFI).  

Data from all array ports at a particular sampling time were ignored in further analysis when a sample in a single port at that time was judged to be an outlier.  Algorithm 1 detected positive outliers by applying an iterative normality test to each array port's total-power time series.  This test compared the sample skewness $g_1$ and sample kurtosis $g_2$ statistics to the respective values of 0 and 3 expected for a Gaussian distribution.  

The rational for this normality test is that we expect the 2~ms resolution total-power time series for the ``hot'' load and ``cold'' sky signals to have a near-Gaussian distribution.  Further, we expect that most potential RFI signals do not have Gaussian distributed total power.  Such use of higher order statistics to detect RFI has been surveyed by \shortcite{Fridman2001}. 

\begin{algorithm}
\label{alg:outliers}
\caption{Detecting outliers in total-power time series of a single port.}
\begin{algorithmic}[1]
\For{$i = 1 \to M$ array ports} 
  \State calculate $g_1$ and $g_2$ for port power time series
  \While{$|g_1| > 0.51 $ \and $|g_2 - 3| > 1.3 $ }
    \State remove sample with largest magnitude
  \EndWhile
\EndFor
\end{algorithmic}
\end{algorithm}



The thresholds at step 3 for limiting excess skewness and kurtosis above their expected values for normality were manually tuned to remove less than 1\% of data from time series judged to contain no RFI on visual inspection.  In the future we could generate a kurtosis threshold for RFI based on a desired false-trigger rate by applying the more rigorously derived spectral kurtosis estimator and associated statistical analysis of \shortcite{Nita2007} and \shortcite{Nita2010a}.

RFI strongly affected measurements at 0.8~GHz, 0.9~GHz and 1.1~GHz at which 6\%, 35\% and 22\% of data were discarded respectively.  Less than 1\% of data were discarded at all other frequencies and there were numerous 0.5~s intervals at particular frequencies where no data was discarded at all.  This highlights an advantage of Algorithm 1: that it will not discard any data that are consistent with a Gaussian distribution.  Thresholding the data at 2.58$\sigma$ would have resulted in typically discarding 1\% of data, in all measurement intervals, that were consistent with a Gaussian distribution. 

We checked for potential bias introduced by Algorithm 1 by comparing overall noise temperature results with and without the application of Algorithm 1.  At all frequencies where less than 10\% of data were discarded by Algorithm 1 (i.e. all except 0.9~GHz and 1.1~GHz) the difference in measured noise temperature with and without Algorithm 1 was less than 0.022~K.  The corresponding difference in uncertainty estimates was less than 0.023~K.  These differences are at least one order of magnitude smaller than the smallest uncertainties in the current measurement procedure (see Figure \ref{fig:error}).

Visual inspection of the 1.1~GHz data suggested that it contained low-duty-cycle transient RFI, likely to be from aviation transponders. This was removed effectively by Algorithm 1.  Inspection of the 0.9~GHz data suggested more continuous RFI, likely to be from mobile telephony.  This was poorly removed by Algorithm 1.  Our experience was consistent with \shortcite{Nita2007} who found that RFI detection based on kurtosis was most effective for low-duty-cycle transient RFI and less effective for continuous RFI.


\section{BEAMFORMED NOISE MEASUREMENT}
We deduce the noise contribution of the array from the Y-factor power ratio between beamformed measurements of the ``hot'' absorber and ``cold'' sky scenes.  We use the notation and unified definitions of efficiencies and system noise temperature for receiving antenna arrays put forward by \cite{Warnick2010}.
\begin{figure*}[!t]
\normalsize
\setcounter{MYtempeqncnt}{\value{equation}}
\setcounter{equation}{12}
\begin{equation}
\label{eqn_dbl_y}
Y = \frac{T_{\text{sys},\text{hot}}}{T_{\text{sys},\text{cold}}} =
  \frac{\eta_\text{rad}(T_{\text{ext},\text{abs}(A)} + T_{\text{ext},\text{sky}(B)} + T_{\text{ext},\text{gnd}}) + T_{\text{loss}} + T_{\text{rec}}}{\eta_\text{rad}(T_{\text{ext},\text{sky}(A)} + T_{\text{ext},\text{sky}(B)} + T_{\text{ext},\text{gnd}}) + T_{\text{loss}} + T_{\text{rec}}}
\end{equation}
\setcounter{equation}{\value{MYtempeqncnt}}
\hrulefill
\vspace*{4pt}
\end{figure*}

Measurements of the receiver-output sample correlation matrix $\mathbf{R}_{xx} $ were made with the array observing a large microwave absorber at ambient temperature giving
\begin{equation}
\begin{split}
  \mathbf{R}_{\text{hot}} &=  \mathbf{R}_{\text{ext},\text{abs}(A)} + \mathbf{R}_{\text{ext},\text{sky}(B)} + \mathbf{R}_{\text{ext},\text{gnd}} \\
  & \ \ \ + \mathbf{R}_{\text{loss}} + \mathbf{R}_{\text{rec}}.
 \end{split}
\end{equation}
Correlation matrix $\mathbf{R}_{\text{ext},\text{abs(A)}} $ measures the thermal noise coupled into the array from the microwave absorber which subtends solid angle $A$ as seen by the array under test. $\mathbf{R}_{\text{ext},\text{sky}(B)} $ measures the stray emission from the sky from solid angle $B$ that is not blocked by the absorber when it is in position and $\mathbf{R}_{\text{ext},\text{gnd}} $ measures stray radiation from the ground which subtends the entire backward hemisphere.  $\mathbf{R}_{\text{loss}} $ is the noise correlation matrix due to ohmic losses in the array and $\mathbf{R}_{\text{rec}} $ is the receiver electronics noise correlation matrix.

A second measurement was made with the array observing the unobstructed radio sky
\begin{equation}
\begin{split}
  \mathbf{R}_{\text{cold}} &=  \mathbf{R}_{\text{ext},\text{sky}(A)} + \mathbf{R}_{\text{ext},\text{sky}(B)} + \mathbf{R}_{\text{ext},\text{gnd}} \\
  & \ \ \ + \mathbf{R}_{\text{loss}} + \mathbf{R}_{\text{rec}}.
 \end{split}
\end{equation}
Beamformed Y-factor was then taken as the ratio of beamformed powers for these two measurements giving
\begin{equation}
  Y = \frac{P_{\text{hot}}}{P_{\text{cold}}} = \frac{\mathbf{w}^H\mathbf{R}_{\text{hot}}\mathbf{w}}{\mathbf{w}^H\mathbf{R}_{\text{cold}}\mathbf{w}}.
  \label{eq:yfact}
\end{equation}
Here $P_{\text{hot}} = G^\text{av}_\text{rec}kBT_{\text{sys,hot}}$ where $G^\text{av}_\text{rec}$ is the available receiver gain, $k$ is Boltzmann's constant, $B$ is the system noise equivalent bandwidth, and $T_{\text{sys,hot}}$ is the beam equivalent system noise temperature of the array under test illuminated by the ``hot'' absorber load.    

When using the definitions of efficiencies and system noise temperature for receiving arrays in \cite{Warnick2010}, the beam equivalent system noise temperature $T_{\text{sys}}$ may be written in the same form as the single-port system noise temperature formula
\begin{equation}
\label{eq:tsyseq}
T_{\text{sys}} = \eta_\text{rad}T_{\text{ext}} + T_{\text{loss}} + T_{\text{rec}}.
\end{equation}
Here $\eta_\text{rad}$ is the beam radiation efficiency, ${T_\text{loss} = (1-\eta_\text{rad})T_{\text{p}}}$ is the beam equivalent noise temperature due to antenna losses, and $T_{\text{p}}$ is the physical temperature of the antenna.  

\cite{Warnick2010} define the beam equivalent system noise temperature $T_{\text{sys}}$ of a receiving antenna array as 
\begin{quote} 
\em ``...the temperature of an isotropic thermal noise environment such that the isotropic noise response is equal to the noise power at the antenna output per unit bandwidth at a specified frequency.''
\end{quote}  

The components of beam equivalent noise temperature due to antenna losses and receiver electronics are both referenced to the antenna ports after antenna losses.  For example, the receiver electronics component of the beam equivalent noise temperature is given by \citep{Warnick2010}
\begin{equation}
T_{\text{rec}} = T_{\text{iso}}\frac{P_{\text{rec}}}{P_{\text{t},\text{iso}}}.
\end{equation}
Here we have normalised by the beam isotropic noise response ${P_{\text{t},\text{iso}}=\mathbf{w}^H\mathbf{R}_{\text{t},\text{iso}}\mathbf{w}}$ which is the beamformed power response of the array to an isotropic thermal noise environment with brightness temperature $T_{\text{iso}}$ when the array itself is in thermal equilibrium at temperature $T_{\text{iso}}$.  Under these conditions $\mathbf{R}_{\text{t},\text{iso}}=\mathbf{R}_{\text{ext},\text{iso}}+\mathbf{R}_\text{loss}$.  

The external contributions from the absorber load, radio sky, and ground are referenced to an antenna temperature before losses, that is ``to the sky''.  For example, the component of the beam equivalent noise temperature due to sky emission from the region of sky blocked by the absorber load is given by \citep{Warnick2010}
\begin{equation}
T_{\text{ext},\text{sky}(A)} = T_{\text{iso}}\frac{P_{\text{ext},\text{sky(A)}}}{P_{\text{ext},\text{iso}}}
\end{equation}
where we have normalised by the beam isotropic noise response ${P_{\text{ext},\text{iso}}=\mathbf{w}^H\mathbf{R}_{\text{ext},\text{iso}}\mathbf{w}}$ before losses.  The pre and post-loss reference planes are referred to each other via the beam radiation efficiency \citep{Warnick2010}
\begin{equation}
\eta_\text{rad} = \frac{P_{\text{ext},\text{iso}}}{P_{\text{t},\text{iso}}} = \frac{P_{\text{ext},\text{iso}}}{P_{\text{ext},\text{iso}}+P_\text{loss}}.
\end{equation}
Combining all of these definitions allows \eqref{eq:yfact} to be rewritten as \eqref{eqn_dbl_y} at the top of this page.\addtocounter{equation}{1}

We define a measurable partial beam equivalent noise temperature 
\begin{equation}
\label{eq:tpart}
{T_\text{n} = \eta_\text{rad}(T_{\text{ext},\text{sky}(B)} + T_{\text{ext},\text{gnd}}) + T_{\text{loss}} + T_{\text{rec}}} 
\end{equation}
 that includes external noise from the sky solid angle $B$ that is not blocked by the absorber and from the ground, and internal noise from antenna losses and receiver electronics.  This is essentially $T_\text{sys}$ less the external sky-noise  $T_{\text{ext},\text{sky}(A)}$ from the solid angle $A$ blocked by the absorber.  This is a step towards the receiver engineer's goal of isolating $T_{\text{loss}}$ and $T_{\text{rec}}$, which are the basic receiver noise performance parameters that should be measured to validate the array design.  
 

We reference the partial beam equivalent noise temperature $T_\text{n}$ ``to the sky'' by dividing through by the beam radiation efficiency $\eta_\text{rad}$.  The sky-referenced partial beam equivalent noise temperature $\hat{T}_\text{n}$ is a quantity that can be determined by inverting \eqref{eqn_dbl_y} at the top of the previous page to give
\begin{equation}
\begin{split}
  \hat{T}_{\text{n}} = \frac{T_{\text{n}}}{\eta_{\text{rad}}} &=  \frac{\alpha T_\text{abs}-YT_{\text{ext},\text{sky}(A)}}{Y-1}.
  \end{split}
  \label{eq:noisetemp}
\end{equation}
Here we have made the substitution  ${T_{\text{ext},\text{abs}(A)} = \alpha T_\text{abs}}$ where $\alpha$ is a beam efficiency factor indicating how well the absorber load fills the beamformed beam and  $T_\text{abs}$ is the physical temperature of the absorber.  We calculate $\alpha$ from the array pattern and absorber geometry in \S\ref{sec:alpha}.  We calculate $T_{\text{ext},\text{sky}(A)}$ from well-established models of the radio sky brightness in \S\ref{sec:sky}.


The ideal case of an infinite absorber $\alpha=1$, zero sky emission ${T_{\text{ext},\text{sky}}=0}$~K, and fixed ambient temperature ${T_\text{abs}=295}$~K reduces \eqref{eq:noisetemp} to
\begin{equation}
  \tilde{T}_\text{n} = \frac{295}{Y-1}.
  \label{eq:trough}
\end{equation}
We have often used \eqref{eq:trough} when order 10~K relative accuracy is acceptable for initial comparison of arrays with identical geometry and test configuration.  When order 1~K absolute accuracy is desired, we use \eqref{eq:noisetemp}.  This is equivalent to making the following systematic corrections to \eqref{eq:trough} 
\begin{equation}
  \hat{T}_{\text{n}} = \frac{\alpha T_\text{abs}}{295} \tilde{T}_\text{n} -  \frac{Y}{Y-1}T_{\text{ext},\text{sky}(A)}.
  \label{eq:troughcorrected}
\end{equation}

Of interest to astronomers wishing to use the array as an aperture-array is the system temperature ${T_\text{sys,cold}}$ when the array observes the unobstructed radio sky.  This is given by
\begin{equation}\label{eq:tsys}
\hat{T}_{\text{sys}} = \frac{T_{\text{sys}}}{\eta_\text{rad}} = \frac{\alpha T_{\text{abs}} - T_{\text{ext},\text{sky}(A)}}{Y-1}.
\end{equation}
The beam equivalent receiver sensitivity can be expressed as \citep{Warnick2008, Warnick2010}
\begin{equation}\label{eq:AeTsys}
\frac{A_\text{e}}{T_\text{sys}} = \frac{\eta_\text{ap}\eta_\text{rad}A_\text{p}}{T_\text{sys}} = \frac{\eta_\text{ap}A_\text{p}}{\hat{T}_\text{sys}} 
\end{equation}
where $A_\text{e}$ is the beam effective area, $\eta_\text{ap}$ is the aperture efficiency, and $A_\text{p}$ is the physical area of the antenna array projected in a plane transverse to the signal arrival direction.

\section{RESULTS}
\begin{figure}
\begin{center}
\includegraphics[width=\columnwidth, angle=0]{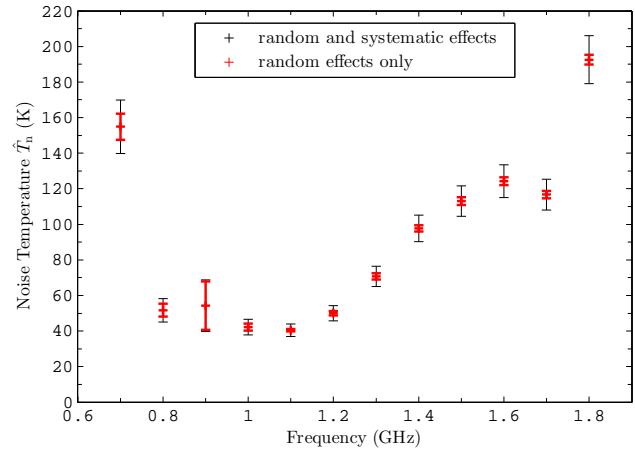}
\caption{Partial beam equivalent noise temperature referenced to the sky ${\hat{T}_\text{n} = T_{\text{ext},\text{sky}(B)} + T_{\text{ext},\text{gnd}} + (T_{\text{loss}} + T_{\text{rec}})/\eta_\text{rad}} $ of the ${5\times4}$ connected-element ``chequerboard'' array.  Maximum $S/N$ weights for a beam directed to zenith were used. Thick, red error bars show uncertainty due to random effects only.  Longer, thin, black error bars show uncertainty due to both random and systematic effects.  The intervals defined by the error bars are believed to contain the unknown values of  ${\hat{T}_\text{n}}$ with a level of confidence of approximately 68 percent.  \label{fig:beamformed-trx}}
\end{center}
\end{figure}

Figure \ref{fig:beamformed-trx} presents the partial beam equivalent noise temperature referenced to the sky ${\hat{T}_\text{n}={T_\text{n}/\eta_\text{rad}}}$ for the prototype ${5\times4}$ array with error bars showing combined standard uncertainty $u_\text{c}(\hat{T}_\text{n})$ (i.e. estimated standard deviation in $\hat{T}_\text{n}$).  Since it can be assumed that the possible estimated values of ${\hat{T}_\text{n}}$ are approximately normally distributed with approximate standard deviation $u_\text{c}(\hat{T}_\text{n})$, the unknown value of ${\hat{T}_\text{n}}$ is believed to lie in the interval ${\hat{T}_\text{n} \pm u_\text{c}(\hat{T}_\text{n})}$ with a level of confidence of approximately 68 percent.  The uncertainty analysis is presented in \S\ref{sec:error} and follows the framework of \shortcite{Taylor1994}.   It applies standard methods for propagating uncertainty in linearly-combined variables to the first-order Taylor-series expansion of \eqref{eq:noisetemp}.    

\begin{figure}
\begin{center}
\includegraphics[width=\columnwidth, angle=0]{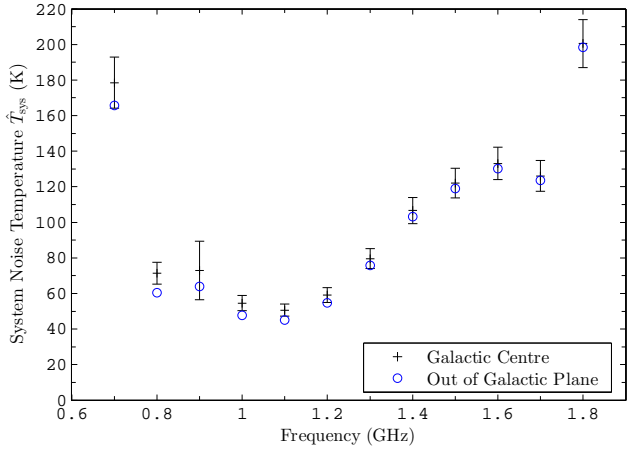}
\caption{Beam equivalent system noise temperature ${\hat{T}_\text{sys} = T_{\text{ext},\text{sky}(A)} + T_{\text{ext},\text{sky}(B)} + T_{\text{ext},\text{gnd}} + (T_{\text{loss}} + T_{\text{rec}})/\eta_\text{rad}} $ of the ${5\times4}$ connected-element ``chequerboard'' array referenced to the sky.  Maximum $S/N$ weights for a beam directed to zenith were used. The data with error bars show the system noise temperature for the measurement configuration of this paper where the array observed the galactic centre.  The intervals defined by the error bars are believed to contain the unknown values of  ${\hat{T}_\text{sys}}$ with a level of confidence of approximately 68 percent.  The circles without error bars show an estimate of the system noise temperature for the array observing out of the galactic plane towards the coldest region of radio sky that transits at the zenith at Parkes (at 3:51~LST).  For clarity of presentation, error bars are not plotted for this second series although they will be very close to a scaled copy of the error bars for the measurement towards the galactic centre. \label{fig:beamformed-tsys}}
\end{center}
\end{figure}

The thick red error bars show a combined standard uncertainty that only includes components of uncertainty arising from random effects.  These are uncertainties $u(P_\text{hot})$ and $u(P_\text{cold})$ and estimated covariance $u(P_\text{hot},P_\text{cold})$ in measurements of the ``hot'' and ``cold'' beamformed powers, and uncertainty $u(T_\text{abs})$ in measurements of the physical temperature of the absorber.  These uncertainties were estimated via statistical methods and are therefore Type A evaluations of uncertainty in the framework of \shortcite{Taylor1994}.       

The thin black error bars show the combined standard uncertainty $u_\text{c}(\hat{T}_\text{n})$ that includes components of uncertainty arising from both random and systematic effects.  The systematic effects included uncertainty in the absorber illumination efficiency $u(\alpha)$ and uncertainty in the beam equivalent external noise temperature due to the radio sky $u(T_{\text{ext},\text{sky}(A)})$ over solid angle $A$ that is blocked by the absorber.  Both of these uncertainties are functions of the beamformed antenna pattern.  They are evaluated via assessments of the range of plausible beam patterns defined by the uniform and optimised weights discussed in \S\ref{sec:syscorr}.  These assessments are Type B (non-statistical) evaluations of uncertainty according to \cite{Taylor1994}.  

The dominant component of uncertainty was the systematic effect characterised by $u(\alpha)$.  This arises from the fact that the beamformed antenna pattern is not measured and so is estimated from theory.  Uncertainty due to random effects was dominated at most frequencies by the contribution of $u(P_{cold})$.  At most frequencies $u(P_{cold})$ characterised noise in measured beamformed power associated with the beam equivalent system temperature. This could be reduced by increasing measurement bandwidths and/or integration times.  However, external RFI was the dominant effect contributing to $u(P_{cold})$ and therefore uncertainty due to random effects at 0.9~GHz. 

For the results in Figure \ref{fig:beamformed-trx} we estimated $T_{\text{ext},\text{sky}(A)}$ using weights with uniform amplitudes and phases that are conjugate matched to the expected spherical wave from the reference radiator.  These same weights are used in \S\ref{sec:alpha} to estimate the lower plausible limit of $\alpha$.  Under the approximate assumption of a direction independent sky brightness, $T_{\text{ext},\text{sky}(A)}$ will be directly proportional to $\alpha$.  Therefore we expect that the uniform amplitude weights should yield an approximate lower bound for $T_{\text{ext},\text{sky}(A)}$.  This should result in a conservative overestimate of $\hat{T}_\text{n}$ via \eqref{eq:noisetemp}.

Figure \ref{fig:beamformed-tsys} shows the beam equivalent system noise temperature referenced to the sky ${\hat{T}_\text{sys}={T_\text{sys}/\eta_\text{rad}}}$.  This is a key factor that determines the receiver sensitivity for an observation towards a particular part of the sky via \eqref{eq:AeTsys}.  It is a property of both the receiver and the receiver's orientation with respect to the sky and surrounding environment.  This is in contrast to ${\hat{T}_\text{n}}$ which is controlled to be as close as practical to a property of the receiver in isolation.

The error bars in Figure \ref{fig:beamformed-tsys} show combined standard uncertainty $u_\text{c}(\hat{T}_\text{sys})$.  A second trace (blue circles) shows the expected reduction in ${\hat{T}_\text{sys}}$ if one of the coldest regions of the sky were used for the ``cold'' scene instead of the hotter galactic centre that was used in this work.  The value of ${\hat{T}_\text{n}}$, on the other hand, is significantly less dependent on the region of sky used as a reference.  Although it becomes clear in \S\ref{sec:error} that using a cold region of sky would reduce uncertainty in ${\hat{T}_\text{n}}$.


\section{SYSTEMATIC CORRECTIONS}
\label{sec:syscorr}
\subsection{Absorber Illumination Efficiency}
\label{sec:alpha}
We define the absorber illumination efficiency $\alpha$ as a dimensionless metric of the beamformed antenna pattern $D(\theta,\phi) $ according to
\begin{equation}
  \alpha = \frac{\int_{A} D(\theta,\phi)d\Omega}{\int_{4\pi}D(\theta,\phi)d\Omega}.
  \label{eq:beam-efficiency}
\end{equation}
This follows the beam efficiency definition of \shortcite{Nash1964} but with the numerator evaluated over the solid angle subtended by the absorber load $A$ instead of the solid angle of the main beam.  This is equivalent to evaluating the solid-beam efficiency, defined in the IEEE Standard Definitions of Terms for Antennas (IEEE Std 145-1993), for solid angle $A$ but ignoring antenna losses.  

\begin{figure}
\begin{center}
\includegraphics[width=\columnwidth, angle=0]{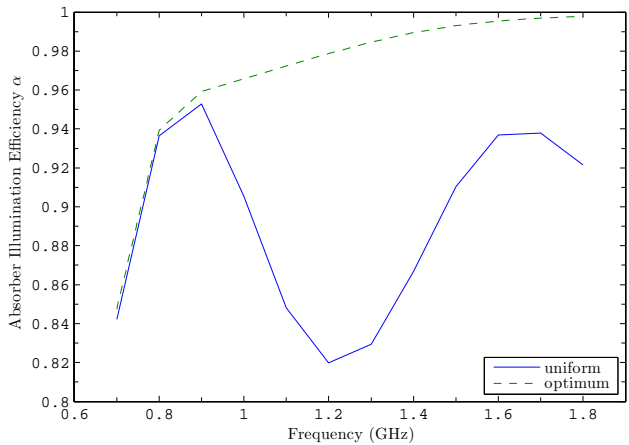}
\caption{Absorber beam illumination efficiency $\alpha$  for uniform amplitude weights and weights with amplitude taper optimised to maximise $\alpha$.  Both sets of weights are conjugate phase matched to the expected spherical wavefront from the reference radiator.  }\label{fig:efficiency}
\end{center}
\end{figure}

Figure \ref{fig:efficiency} shows values of $\alpha$ calculated for array patterns formed by two different weightings of 40 isotropic elements arranged with the same geometry as the ${5 \times 4}$ prototype.  The weight phases were conjugate matched to a supposed spherical wavefront emanating from the reference antenna used to constrain beam pointing.  Weight amplitudes were assigned according to two different methods: uniform amplitudes and an optimised amplitude taper.  These choices are thought to encompass the plausible range of amplitude tapers imposed by the maximum $S/N$ weights of \eqref{eq:maxsnrweights} with direction and polarisation constrained by the reference antenna measurement.  

We explored a third amplitude taper function matched to the expected illumination from the LPDA reference antenna according to pattern measurements provided by its manufacturer.  The LPDA $\alpha$ result was not plotted as it was within 1\% of that obtained by the uniform amplitude weights. 

Based on the range of $\alpha$ exhibited in Figure \ref{fig:efficiency} we assessed that the value of $\alpha$ was highly likely (near 100\% probability) to lie in the range $\alpha = 0.9 \pm 0.1$.  Uncertainty in $\alpha$ was modelled by a uniform distribution over this range.  We divided the half-range of this distribution of 0.1 by $\sqrt{3}$, according to \cite{Taylor1994}, to estimate the standard uncertainty in the absorber illumination efficiency $u(\alpha)=0.0577$. 

The uniform-weight pattern is easy to calculate and we expect it to give the narrowest main beam but with high side lobes. This should perform well at lower frequencies where the ${5\times4}$ array is too small to form a main beam that falls entirely within the area blocked by the absorber load.  In fact, the uniform-weight $\alpha$ turned out to be consistent with the optimised-taper $\alpha$ below 1~GHz.


The optimised taper was calculated by parameterising an amplitude taper for an ideal boresight beam with the following taper function that is separable on $x$ and $y$ coordinates \citep{Nash1964}
\begin{equation}\label{eq:taper}
  \begin{split}
  |w| = \ \ &\left[K_x + (1-K_x)(1-(2x/L_{x})^2)^{n_x}\right]\\
  \times &\left[K_y + (1-K_y)(1-(2y/L_{y})^2)^{n_y}\right].
  \end{split}
\end{equation}
Here $L_x$ is the linear size of the array along the $x$-axis.  Parameters $K_x$ and $n_x$ determine the shape of the taper function factor that separates along the $x$-axis. 



We tried two constrained optimisation techniques to find the parameters of \eqref{eq:taper} that lead to an array pattern that maximised $\alpha $ as calculated by \eqref{eq:beam-efficiency}, subject to the constraints ${0.1 < K < 0.999}$ and ${0.5 < n < 4}$.  Both the SNOPT implementation \citep{Gill2013, Gill2005} of sequential quadratic programming (SQP) optimisation and Standard Particle Swarm Optimisation SPSO-2011 \citep{Clerc2012,Zambrano-Bigiarini2013} yielded the same optimal value of $\alpha$ to within 0.04\%.

We calculated the array pattern assuming isotropic elements (array factor) as we wanted a measurement and analysis method that does not require special knowledge of the array design beyond its geometry.  This allows measurement and comparison of arrays provided as complete ``black-box'' systems.  Our technique will become even more accurate for larger arrays, such as the ASKAP 188-port PAFs, that can form narrower beams with lower side lobes and therefore more efficiently illuminate the load. 

A better estimate of the partial beam equivalent noise temperature may be formed by including simulated or measured element patterns, but this is beyond the scope of the current work.  Our technique is fair for the current array which has element patterns with relatively low gain.    Alternatively, we could force $\alpha$ closer to unity by employing a ground shield to reflect as much of the array pattern as possible onto the load, by reducing the vertical spacing between the load and the array under test, or by using a larger load.  We are building a ground shield and a larger load for future experiments.

\subsection{Sky Brightness}\label{sec:sky}
\begin{figure}
\begin{center}
\includegraphics[width=\columnwidth, angle=0]{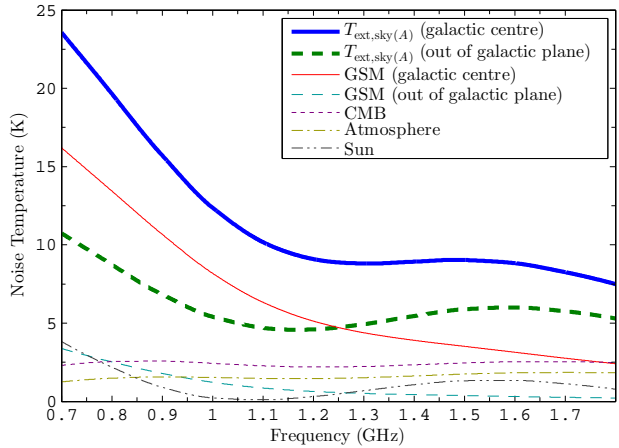}
\caption{Breakdown of contributions to $T_{\text{ext},\text{sky}(A)}$, which is the beam equivalent external noise temperature due to radio emission from the area of sky blocked by the absorber load.  The Global Sky Model (GSM) contribution is calculated at 17:39~LST, near transit of the galactic centre, when the measurements for this paper were made.  It is also calculated at 03:51~LST when zenith observations from latitudes near $30^\circ$S point out of the galactic plane towards one of the coldest patches of radio sky as deduced by measured and modelled drift-scans in \S7 of \shortcite{Chippendale2009}.  The thick lines show total $T_{\text{ext},\text{sky}(A)}$ for these two limiting observation epochs.  The thin lines show the breakdown of these totals into contributions from the GSM, cosmic microwave background (CMB), atmosphere, and Sun.  The curves for the diffuse backgrounds (i.e. all but the Sun curve) would be directly proportional to $\alpha$ if the sky brightness were direction independent.  }\label{fig:tskybreakdown}
\end{center}
\end{figure}

Calculating $\hat{T}_\text{n}$ via \eqref{eq:noisetemp} or $\hat{T}_\text{sys}$ via \eqref{eq:tsys} requires an estimate of $T_{\text{ext},\text{sky}(A)}$.  This is the component of pre-loss beam equivalent noise temperature  due to radio emission from the region of sky $A$ blocked by the absorber when in position for the ``hot'' measurement. Figure \ref{fig:tskybreakdown} shows a break-down of contributions to $T_{\text{ext},\text{sky}(A)}$ for measurements made from the Parkes Test Facility towards both the hottest and coolest regions of the radio sky observed during zenith pointing drift scans.  These have been calculated with the same uniform amplitude weights used to estimate $\alpha$ in \S\ref{sec:alpha}.  

We estimated $T_{\text{ext},\text{sky}(A)}$ by convolving the sky brightness $T_\text{bsky}(\theta,\phi)$ with the beamformed antenna pattern $D(\theta,\phi)$ to give 
\begin{equation}\label{eq:tasky}
  T_{\text{ext},\text{sky}(A)} = \frac{\int_{A}T_\text{bsky}(\theta,\phi)D(\theta,\phi)d\Omega}{\int_{4\pi}D(\theta,\phi)d\Omega}.
\end{equation}
The numerator is evaluated over the solid angle $A$ subtended by the absorber load when in position over the array under test as shown in Figure \ref{fig:hot-load}.  The denominator normalises by the beam solid angle.


The sky brightness is modelled as background radio sky brightness $T_\text{b0} $, attenuated by a dry atmosphere with: air mass $X(\theta)$ as a function of zenith angle $\theta $,  transmissivity  $e^{-\tau X(\theta)} $, and atmospheric noise emission represented by an equivalent physical temperature $T_\text{atm} $
\begin{equation}
\begin{split}
  T_\text{bsky}(\theta,\phi) = \ \ &T_\text{b0}(\theta,\phi)e^{-\tau X(\theta)} \\
                          + &(1 - e^{-\tau X(\theta)})T_\text{atm}. 
\end{split}
\end{equation}

Background sky brightness $T_\text{b0} $ was estimated at each measurement frequency using the global radio sky model (GSM) of \shortcite{deOliveira-Costa2008} plus an isotropic Cosmological Microwave Background (CMB) contribution of 2.725~K \citep{Fixsen2009}.  The integral in \eqref{eq:tasky} was evaluated with 0.5$^\circ$ resolution in $\theta$ and $\phi$, which exceeds the 1$^\circ$ resolution of the GSM evaluated at the frequencies of interest with principal component amplitudes locked to the the 408~MHz map of \shortcite{Haslam1982}.  

The Sun's contribution is considered by adding a single pixel to $T_\text{b0}(\theta,\phi)$ with brightness temperature $\bar{T}_\Sun\bar{\Omega}_\Sun / \Omega_\text{pixel}$.  Here $\bar{T}_\Sun$ is the sum of the steady state component of solar emission plus the mean of the slowly changing component, all normalised to the mean visible solid angle of the Sun $\bar{\Omega}_\Sun = 0.22~\text{deg}^2$.  The spectrum of $\bar{T}_\Sun$ was interpolated by fitting a power law to the single-frequency values tabulated by \cite{Kuzmin1966} and reproduced here in Table~\ref{sunflux} for convenience.

\begin{table}
\caption{Intensity of Solar Radio Emission \citep{Kuzmin1966}.}
\begin{center}
\begin{tabular}{@{}cc@{}}
\hline\hline
Frequency & Mean Brightness Temperature$^\text{a}$ \\

$f$ (GHz)       &  $\bar{T}_\Sun$ (K)\\ 
\hline
 0.6  & $\frac{6\times10^5}{3\times10^5}^{\text{b}}$  \\ 
 1.2  & $\frac{2\times10^5}{1\times10^5}$  \\ 
 3.0  & $\frac{8\times10^4}{4\times10^4}$ \\
\hline\hline
\end{tabular}
\end{center}
\begin{footnotesize}
$^\text{a}$ The mean brightness temperature is referenced to the mean visible solid angle of the Sun ${\bar{\Omega}_\Sun=0.22 \ \text{deg}^2}$.  This table gives the constant component and the mean value of the slowly varying component.\\
$^\text{b}$ The numerator corresponds to a period of maximum solar activity and the denominator corresponds to a period of minimum solar activity.

\end{footnotesize}
\label{sunflux}
\end{table}

 Dry atmosphere transmissivity $e^{-\tau} $ at zenith was calculated according to Annex 2 of ITU Recommendation ITU-R P.676-9; typical equivalent physical temperature of the atmosphere $T_\text{atm} = 275$~K was taken from ITU-R P.372-10; and the air mass versus zenith angle $X(\theta) $ model fit of \shortcite{Young1994} was used. 

\section{STRAY EXTERNAL NOISE}
The ``stray'' beam equivalent external noise can be broken into sky and ground components.  The stray-sky noise $T_{\text{ext},\text{sky}(B)}$ may be estimated via the same method as $T_{\text{ext},\text{sky}(A)}$ in \eqref{eq:tasky}, but evaluating the integral in the numerator over the area of sky not blocked by the absorber which we label solid angle $B$.  We assumed $T_{\text{ext},\text{sky}(B)}$ was unchanged between ``hot'' and ``cold'' measurements, and therefore neglected scattering from the sparse metal frame that supports the absorber.

\begin{figure}
\begin{center}
\includegraphics[width=\columnwidth, angle=0]{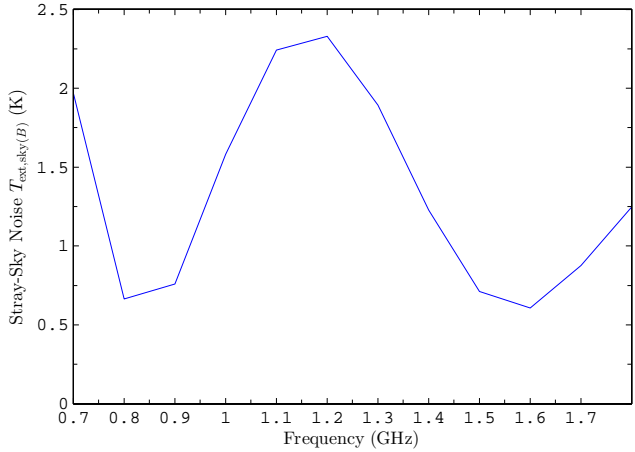}
\caption{Antenna temperature component due to stray-sky radiation calculated via \eqref{eq:tasky}, but evaluating the integral in the numerator over the area of sky $B$ not blocked by the absorber.  This would be directly proportional to ${(1-\alpha)}$ if the sky brightness were direction independent.}\label{fig:tstray}
\end{center}
\end{figure}

Figure \ref{fig:tstray} shows the resulting $T_{\text{ext},\text{sky}(B)}$ estimate for the experiment presented in this paper. It will be highest when the galactic centre is almost but not quite blocked by the absorber.  It will be lowest when the galactic centre is near/below the horizon or completely blocked by the absorber.

The stray-ground radiation  $T_{\text{ext},\text{gnd}}$ can also be estimated via \eqref{eq:tasky}, but evaluating the integral in the numerator over the backward hemisphere and substituting the sky-brightness model with a ground-brightness model $T_\text{g}(\theta,\phi)$.  An order of magnitude estimate of $T_{\text{ext},\text{gnd}}$ may be made by assuming that the ground brightness takes on the direction independent value of $T_\text{g}$.  We would then estimate $T_{\text{ext},\text{gnd}} = (1-e_\text{f})T_\text{g}$.  Here $e_\text{f}$ is the forward efficiency of the beamformed antenna pattern.  This may be calculated via \eqref{eq:beam-efficiency}, but with the numerator evaluated over the full forward hemisphere instead of just the solid angle blocked by the absorber.

Stray radiation could be measured by making beamformed Y-factor measurements with and without a ground-shield.  As mentioned above, such a ground shield is being manufactured for ongoing noise measurements of array receivers at the Parkes Test Facility.  

\section{MEASUREMENT UNCERTAINTY}\label{sec:error}
Uncertainty in the measured noise temperature was estimated via the framework of \shortcite{Taylor1994}.  The combined standard error $u_\text{c}(\hat{T}_\text{n})$ is an estimate of the standard deviation in $\hat{T}_\text{n}$.  This estimate is made via the linear combination of uncertainties in the first order Taylor series expansion of \eqref{eq:noisetemp}.  This gives 
\begin{equation}
\begin{split}
  u_\text{c}^2(\hat{T}_\text{n}) = 
  &\ \ \left(\frac{ \partial \hat{T}_\text{n} }{ \partial P_\text{hot} }\right)^2 u^2(P_\text{hot}) \\ &\!\!+  \left(\frac{ \partial \hat{T}_\text{n} }{ \partial P_\text{cold} }\right)^2 u^2(P_\text{cold})\\
&\!\!+ 2\left|\frac{\partial \hat{T}_\text{n} }{\partial P_\text{hot} }\frac{\partial \hat{T}_\text{n} }{\partial P_\text{cold} }\right| u\left(P_\text{hot}, P_\text{cold} \right) \\
&\!\!+ \left(\frac{ \partial \hat{T}_\text{n}}{\partial \alpha}\right)^2 u^2( \alpha )   + \left(\frac{ \partial \hat{T}_\text{n}}{\partial T_\text{abs}}\right)^2 u^2( T_\text{abs})\\
&\!\!+ \left(\frac{ \partial \hat{T}_\text{n}}{\partial T_{\text{ext},\text{sky}(A)}}\right)^2 u^2( T_{\text{ext},\text{sky}(A)}) 
\end{split}
\end{equation}
where $u(x)$ is an estimate of the standard deviation associated with input estimate $x$ and $u(x,y)$ is an estimate of the covariance associated with input estimates $x$ and $y$.  Evaluation of the partial derivatives of \eqref{eq:noisetemp} and substitution of \eqref{eq:yfact} and \eqref{eq:tsys} yields the square of the relative combined standard uncertainty as a function of input estimates 
\begin{equation}
  \begin{split}\label{eq:error}
\!\!\!\!\!\!\! \left( \frac{ u_c(\hat{T}_\text{n} )}{\hat{T}_\text{n} }\right)^2 &= 
   \Bigg\{ \left(\frac{\hat{T}_{sys}}{\hat{T}_n}\right)^2 \left[ \left(\frac{u(P_\text{hot})}{P_\text{hot}}\right)^2 \right.\\
  & \ \ \ \ \ \ \ \ \ \ \ \ \ \ \ \ \ + \left(\frac{u(P_\text{cold})}{P_\text{cold}}\right)^2  \\
  & \ \ \ \ \ \ \ \ \ \ \ \ \ \ \ \ \ + \left.\frac{2u\left(P_\text{hot},P_\text{cold}\right)}{P_\text{hot}P_\text{cold}}\right] \\
&\!\!\!\!\!\!\!\!\!\!\!\!\!\!\!\!\!\!\!\!\!\!\!\!\!\!\!\!\!+ \left(\frac{\alpha T_\text{abs}}{Y\hat{T}_\text{n}}\right)^2 \left[\left( \frac{ u(\alpha )}{ { \alpha } } \right)^2 + \left( \frac{ u( T_\text{abs} )}{ { T_\text{abs}} } \right)^2 \right] \\
&\!\!\!\!\!\!\!\!\!\!\!\!\!\!\!\!\!\!\!\!\!\!\!\!\!\!\!\!\!+ \left(\frac{T_{\text{ext},\text{sky}(A)}}{\hat{T}_\text{n}}\right)^2 \left( \frac{ u( T_{\text{ext},\text{sky}(A)} )}{{ T_{\text{ext},\text{sky}(A)} } } \right)^2 \Bigg\}\left(\frac{Y}{Y-1}\right)^2\!.
\end{split}
\end{equation}
Applying the same uncertainty analysis to $\hat{T}_\text{sys}$ as defined by \eqref{eq:tsys} yields
\begin{equation}
  \begin{split}\label{eq:errortsys}
\!\!\!\!\!\!\! \left( \frac{ u_c(\hat{T}_\text{sys} )}{\hat{T}_\text{sys} }\right)^2 &= 
   \Bigg\{ \left[ \left(\frac{u(P_\text{hot})}{P_\text{hot}}\right)^2  + \left(\frac{u(P_\text{cold})}{P_\text{cold}}\right)^2 \right.\\
  & \ \ \ \ \ \ \ \ \ \ \ \ \ \ \ \ \ + \left.\frac{2u\left(P_\text{hot},P_\text{cold}\right)}{P_\text{hot}P_\text{cold}}\right] \\
&\!\!\!\!\!\!\!\!\!\!\!\!\!\!\!\!\!\!\!\!\!\!\!\!\!\!\!\!\!\!\!\!+ \left(\frac{\alpha T_\text{abs}}{Y\hat{T}_\text{sys}}\right)^2 \left[\left( \frac{ u(\alpha )}{ { \alpha } } \right)^2 + \left( \frac{ u( T_\text{abs} )}{ { T_\text{abs}} } \right)^2 \right] \\
&\!\!\!\!\!\!\!\!\!\!\!\!\!\!\!\!\!\!\!\!\!\!\!\!\!\!\!\!\!\!\!\!+ \left(\frac{T_{\text{ext},\text{sky}(A)}}{Y\hat{T}_\text{sys}}\right)^2 \left( \frac{ u( T_{\text{ext},\text{sky}(A)} )}{{ T_{\text{ext},\text{sky}(A)} } } \right)^2 \Bigg\}\left(\frac{Y}{Y-1}\right)^2\!.
\end{split}
\end{equation}
In both \eqref{eq:error} and \eqref{eq:errortsys}, the first term's dependence on $u(P_{\text{hot}})$ and $u(P_\text{cold})$ suggests that beamformed power measurements should be made with adequate integration time and/or measurement bandwidth to reduce measurement variance via averaging. The first term's dependence on $u(P_\text{hot},P_\text{cold})$ highlights the necessity to minimise gain and/or noise performance drift between ``hot'' and ``cold'' measurements. The second term highlights the importance of knowing the absorber-illumination efficiency $\alpha$ and accurately measuring the ambient temperature of the absorber.  Good knowledge of the beam pattern of the array under test is required to accurately estimate $\alpha$.  The third term highlights the importance of estimating the beamformed antenna temperature which is a function of the background sky brightness and the beamformed antenna pattern.  Comparing \eqref{eq:error} and \eqref{eq:errortsys} shows that ${u( T_{\text{ext},\text{sky}(A)} )}$ contributes less uncertainty to ${\hat{T}_\text{sys}}$ than to ${\hat{T}_\text{n}}$, particularly for low-noise arrays under test with high Y-factors. 

\begin{figure}
\begin{center}
\includegraphics[width=\columnwidth, angle=0]{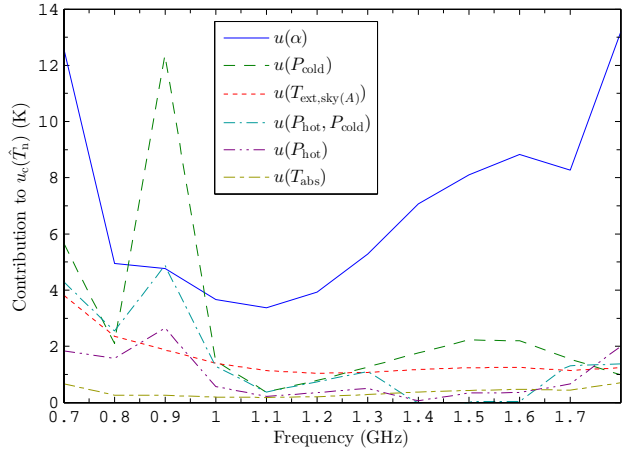}
\caption{Breakdown of contributions to combined uncertainty $u_\text{c}(\hat{T}_\text{n})$ in partial beam equivalent noise noise temperature due to each input uncertainty in \eqref{eq:error}. }\label{fig:error}
\end{center}
\end{figure}

Figure \ref{fig:error} shows the contribution of each input uncertainty, via \eqref{eq:error}, to the combined standard uncertainty $u_\text{c}(\hat{T}_\text{n})$ in partial beam equivalent noise temperature.  This shows that the combined uncertainty is largely dominated by uncertainty $u(\alpha)$ in the illumination of the absorber, which is in turn due to uncertainty in the beamformed antenna pattern.

In the above uncertainty analysis we have not included the correlation between $\alpha$ and $T_{\text{ext},\text{sky}(A)}$.  This correlation arises as they are both direct functions of the beamformed antenna pattern.  In the future, we could take advantage of the fact that the stray external noise $T_{\text{ext},\text{sky}(B)}$ is also a function of the antenna pattern and correlated with $\alpha$ and $T_{\text{ext},\text{sky}(A)}$.  Including these correlations in the uncertainty analysis at the same time as subtracting an estimate of $T_{\text{ext},\text{sky}(B)}$ from $\hat{T}_\text{n}$ may lead to some cancellation in uncertainty terms that depend on the antenna pattern.  This could reduce uncertainty at the same time as moving us closer to extracting $T_\text{loss}$ and $T_\text{rec}$ from $\hat{T}_\text{n}$.

\section{CONCLUSION}
We have demonstrated the measurement of partial beam equivalent noise temperature as low as ${\hat{T}_\text{n}=40 \ \text{K}}$ with a combined standard uncertainty (estimated standard deviation) as low as ${u_\text{c}(\hat{T}_\text{n})=4 \ \text{K}}$.  This combined uncertainty was dominated by uncertainty $u(\alpha)$ in the efficiency with which the beamformed array pattern illuminates the absorber. 

The prioritised action-list for reducing uncertainty further is:
\begin{enumerate}
  \item Reducing uncertainty in absorber-illumination efficiency $\alpha$ by:
  \begin{itemize}
    \item increasing the solid angle subtended by the absorber by increasing its size or moving it closer to the array under test, 
    \item adding a ground shield to reflect more of the antenna pattern onto the absorber, or
     \item accurately measuring or modelling the beamformed antenna pattern.
  \end{itemize}
  \item Moving the reference radiator into the far field of the array under test.  
  \item Increasing integration time for the ``cold'' sky measurement.
  \item Reducing uncertainty in beam equivalent noise temperature due to radio emission from the sky $T_{\text{ext},\text{sky}(A)}$ by:
  \begin{itemize}
    \item using the coldest possible region of the sky for the ``cold'' sky measurement,
    \item accurately measuring or modelling the beamformed antenna pattern, and
    \item improving the accuracy of the global sky model.
  \end{itemize}
  
\end{enumerate} 
Addressing items (1) to (3) would reduce the median combined standard uncertainty to just ${u_\text{c}(\hat{T}_\text{n})=2 \ \text{K}}$ over 0.7~GHz to 1.8~GHz.


\section{ONGOING DEVELOPMENT}
After the measurements were made for this paper, the Parkes test facility was upgraded to include a 192-port down-conversion and digital receiver system.  This supports 304~MHz instantaneous bandwidth tunable over 0.7~GHz to 1.8~GHz with 1~MHz spectral resolution.  It is capable of online measurement of the full $192\times192$ correlation matrix and online beamforming for nine simultaneous dual-polarisation beams. 

This was achieved by installing the electronics that are normally found in the pedestal of ASKAP's first six ``BETA'' antennas \citep{Schinckel2011, Bunton2011} into a hut near the test-pad.  This receiver can be connected to test arrays mounted on the aperture-array test pad or at the focus of the nearby 12~m dish via RF coaxial cables in trenches.  The upgraded facility was recently used to verify an enhanced ASKAP LNA and chequerboard array design that has low-noise performance over the full 0.7~GHz to 1.8~GHz band \citep{Shaw2012}.  This  improvement will be included in the ASKAP Design Enhancements (ADE) PAF \citep{Hampson2012}.   The facility was also used to characterise the astronomical performance of a 188-port BETA PAF that is currently installed at the focus of the 12~m dish.

In the future, better estimates of the array noise properties may be obtained by electromagnetic modelling of test setups and the external environmental contribution with or without a shield.  Work is underway to build a ground shield to both reduce $u(\alpha)$ and allow estimation of ``stray'' beam equivalent external noise ${T_\text{ext,sky}(B) + T_\text{ext,gnd}}$.  We are also building a larger absorber load for use at our radio quiet site at the Murchison Radioastronomy Observatory (MRO) where ASKAP is sited.  This radio quiet site will allow more repeatable noise measurements below 1~GHz where the RFI situation at Parkes becomes challenging.



\begin{acknowledgements}
The Australian SKA Pathfinder is part of the Australia Telescope National Facility which is funded by the Commonwealth of Australia for operation as a National Facility managed by CSIRO.  The ``chequerboard'' array and system development has been the work of the ASKAP team.  

The LNAs and downconversion modules for the prototype presented in this paper were designed by Alex Grancea.  John Bunton developed the digital system architecture  which was implemented by Joseph Pathikulangara, Jayasri Joseph, Tim Bateman, Andrew Brown and Dezso Kiraly.  Malte Marquarding, Juan Carlos Guzman, Euan Troup, David Brodrick and Simon Hoyle developed and supported software to operate the system.  Tim Wilson did the mechanical design for the structure supporting the absorber load. The staff at the CSIRO Parkes Observatory, including Brett Preisig, Ian McRobert, Tom Lees and Jon Crocker have supported the testing by developing the infrastructure, including the absorber load, and responding rapidly to diverse needs during measurement campaigns.    

The array measurements at Parkes, leading up to and including the measurements presented in this paper, were conducted by Robert Shaw, Ian McRobert, Tim Bateman, Peter Axtens, Russell Gough, John O'Sullivan and Kjetil Wormnes in addition to the authors.  Preparation work in Sydney has been supported by many colleagues, including Carl Holmesby, Ivan Kekic and Ken Smart.  Russell Gough, Aidan Hotan and John Bunton also provided helpful comments on intermediate drafts. 
\end{acknowledgements}



\end{document}